\documentclass[12pt,a4paper,nohyper]{JHEP}
\usepackage{subeqn}
\usepackage{amssymb}
\usepackage{cite}
\usepackage{epsfig}
\def\unit{\leavevmode\hbox{\small1\kern-3.6pt\normalsize1}}
\def \be{\begin{equation}}
\def \ee{\end{equation}}
\def \bea{\begin{eqnarray}}
\def \eea{\end{eqnarray}}
\def \ben{\begin{enumerate}}
\def \een{\end{enumerate}}
\def \bit{\begin{itemize}}
\def \eit{\end{itemize}}
\def \av#1{\left\langle #1\right\rangle}
\def \eff{\mathrm{eff}}
\def \vckm{V_{\mathrm{CKM}}}
\def \subfermion{\mathrm{fermion}}
\def \subhiggs{\mathrm{Higgs}}
\def \sb{\mathrm{SB}}
\def \susy{\mathrm{SUSY}}
\def \GeV{{\mathrm{GeV}}}

\def \MeV{{\mathrm{MeV}}}
\def \TeV{{\mathrm{TeV}}}
\def \Im{{\mathrm{Im}}\,}
\def \Re{{\mathrm{Re}}\,}

\def \bm{\boldmath}

\def \braket#1#2#3{\langle #1|#2| #3\rangle}

\def \cl#1{{#1\%\ \mathrm{C.L.}}}

\def \cp{\mathrm{CP}}

\def \diag{{\mathrm{diag}}}

\def \ea{{\it et al.}}
\def \eq#1{Eq.~(\ref{#1})}
\def \eqs#1#2{Eqs.~(\ref{#1})--(\ref{#2})}
\def \fig#1{Fig.~\ref{#1}}

\def \heff{H_{\mathrm{eff}}}
\def \hc{\mathrm{H.c.}}
\def \nnu{\nonumber}

\def \ol#1{\overline{#1}}

\def \rf{Ref.~\cite}
\def \rfs{Refs.~\cite}
\def \sec#1{Sec.~\ref{#1}}

\def \b{\beta}
\def \f{\phi}
\def \D{\Delta}
\def \g{\gamma}

\def \d{\delta}
\def \epsi{\epsilon}

\def \l{\lambda}

\def \m{\mu}

\def\21{$SU(2) \ot U(1)$} 
\def\ot{\otimes}

\def\vev#1{\left\langle #1\right\rangle}

\def\bold#1{\setbox0=\hbox{$#1$} 
     \kern-.025em\copy0\kern-\wd0 
     \kern.05em\copy0\kern-\wd0 
     \kern-.025em\raise.0433em\box0 }
\def \C{\tilde{\chi}}
\def \chargino{\tilde{\chi}^{\pm}}

\def \gluino{\tilde{g}}
\def \higgs{H^{\pm}}
\def \neutralino{\tilde{\chi}^0}

\def \squark{\tilde{q}}

\def \squarki#1{\tilde{q}_{#1}}

\def \wino{\tilde{W}}

\def \higgsino{\tilde{H}}
\def\ann#1#2#3{{\it Annu. Rev. Nucl. Part. Sci.\/} {\bf#1} (19#2) #3}
\def\euro#1#2#3{{\it Eur. Phys. J.\/} {\bf C#1} (19#2) #3}
\def\euron#1#2#3{{\it Eur. Phys. J.\/} {\bf C#1} (20#2) #3}
\def\ijmp#1#2#3{{\it Int.~J.~Mod.~Phys.\/}~{\bf A#1} (19#2) #3}
\def\ijmpn#1#2#3{{\it Int.~J.~Mod.~Phys.\/}~{\bf A#1} (20#2) #3}
\def\ibid#1#2#3{\emph{ibid.} {\bf#1} (19#2) #3}
\def\ibidn#1#2#3{\emph{ibid.} {\bf#1} (20#2) #3}

\def\np#1#2#3{{\it Nucl.~Phys.\/}~{\bf B#1} (19#2) #3}
\def\npn#1#2#3{{\it Nucl.~Phys.\/}~{\bf B#1} (20#2) #3}

\def\pl#1#2#3{{\it Phys.~Lett.\/}~{\bf B#1} (19#2) #3}
\def\pln#1#2#3{{\it Phys.~Lett.\/}~{\bf B#1} (20#2) #3}

\def\prd#1#2#3{{\it Phys.~Rev.\/}~{\bf D#1} (19#2) #3}
\def\prdn#1#2#3{{\it Phys.~Rev.\/}~{\bf D#1} (20#2) #3}
\def\prl#1#2#3{{\it Phys.~Rev.~Lett.\/}~{\bf #1} (19#2) #3}
\def\prln#1#2#3{{\it Phys.~Rev.~Lett.\/}~{\bf #1} (20#2) #3}
\def\prp#1#2#3{{\it Phys.~Rep.\/}~{\bf #1} (19#2) #3}
\def\ptp#1#2#3{{\it Prog. Theor.~Phys.\/}~{\bf #1} (19#2) #3}
\def\rmp#1#2#3{{\it Rev. Mod. Phys.\/} {\bf #1} (19#2) #3}
\def\zpc#1#2#3{{\it Z.~Phys.\/}~{\bf C#1} (19#2) #3}
\preprint{CERN-TH/2000-356\\
FISIST/19-00/CFIF\\
TUM-HEP-397/00}   
\title{Spontaneous \bm$\cp$ Violation in the Next-to-Minimal
Supersymmetric Standard Model Revisited}
\author{G.C. Branco,$^{a,b}$ F. Kr\"uger,$^c$ J.C. Rom\~ao,$^b$ and 
A.M.~Teixeira$^b$\\
$^a$Theory Division, CERN, CH-1211 Geneva 23, Switzerland\\
$^b$Centro de F\'\i sica das Interac\c{c}\~{o}es Fundamentais (CFIF),
Departamento de F\'{\i}sica,  Instituto Superior T\'ecnico,  
Av. Rovisco Pais,  1049-001 Lisboa, Portugal\\
$^c$Physik Department, Technische Universit\"at M\"unchen,
D-85748 Garching, Germany\\
E-mail: \email{gbranco@cfif.ist.utl.pt, fkrueger@ph.tum.de,
fromao@alfa.ist.utl.pt, ana@cfif.ist.utl.pt}}
\abstract{We re-examine 
spontaneous $\cp$ violation at the tree level in the context of 
the next-to-minimal supersymmetric standard model (NMSSM) 
with two Higgs doublets and a gauge singlet field. 
We analyse the most general Higgs potential without a discrete 
$Z_3$ symmetry, and derive an upper bound on the mass of the lightest
neutral Higgs boson consistent with present experimental data.  
We investigate, in particular, its dependence on the admixture 
and $\cp$-violating phase of the gauge singlet field, as well as on 
$\tan\b$.
To assess the viability of the spontaneous $\cp$ violation scenario,
we estimate $\epsi_K$ by applying the mass insertion approximation. 
We find that a non-trivial flavour structure in the 
soft-breaking $A$ terms is required to account for the observed 
$\cp$ violation in the neutral kaon sector. Furthermore,  
combining the minimisation conditions for spontaneous CP violation with
the constraints coming from $K^0$--$\bar{K}^0$ mixing, we find that 
the upper bound on the lightest Higgs-boson mass becomes stronger.
We also point out that the electric dipole moments of electron and 
neutron are a serious challenge for SUSY models with spontaneous CP
violation.
}
\begin{document}
\section{Introduction}\label{intro}
The origin of $\cp$ violation remains a fundamental open question in
particle physics. In the standard model (SM), $\cp$ is explicitly broken
at the Lagrangian level through complex Yukawa couplings which lead to
$\cp$ violation in charged weak interactions via the
Cabib\-bo-Ko\-ba\-ya\-shi-Maskawa (CKM) matrix,  $\vckm$ \cite{ckm}. 
Although the CKM mechanism can accommodate the experimental value of $\epsi_K$
(and, in principle, also $\epsi'/\epsi_K$) in the neutral kaon sector, it is 
not clear whether it is the dominant contribution or the only one. 
An important motivation to consider new
sources of $\cp$ violation stems from the fact that within the SM 
the amount of $\cp$ violation may not be sufficient to generate the observed 
baryon asymmetry in the universe \cite{baryogenisis}.

An alternative scenario for the breaking of $\cp$ is to assume that it is
a symmetry of the Lagrangian which is only spontaneously broken
by the vacuum \cite{TDlee}. The purpose of the present paper is to study 
spontaneous 
breaking of $\cp$  at the tree level within the context of
supersymmetry (SUSY).
Although the minimal supersymmetric standard model (MSSM) has two
Higgs doublets, it is well 
known that due to SUSY constraints on the Higgs potential it is not possible to obtain spontaneous 
$\cp$ violation (SCPV) at the tree level. The reason for this is entirely analogous to the 
situation one encounters in non-supersymmetric two-Higgs-doublet models, where SCPV cannot be achieved if a $Z_2$ symmetry is imposed on the Lagrangian in 
order to guarantee 
natural flavour conservation in the Higgs sector
\cite{gustavo:scpv}. The possibility that SCPV in the MSSM might be 
generated through radiative corrections has been explored in 
\rf{mssm:cp:spontaneous}, but this particular scenario has already been
ruled out by experiment as it inevitably leads to the existence of a
very light Higgs boson. Ultimately, this result is a consequence 
of the Georgi-Pais theorem \cite{georgi-pais}. It is therefore of
interest to consider simple extensions of the MSSM such as a model
with at least one gauge singlet field ($N$) besides the two Higgs
doublets ($H_{1,2}$), the so-called next-to-minimal supersymmetric 
standard model 
(NMSSM) \cite{model:nmssm:I,model:nmssm:II}, and to ask if 
one can achieve spontaneous breaking of $\cp$ whilst generating the
observed amount of $\epsi_K$ and having Higgs-boson masses
that are consistent with experimental data 
\cite{susy:cp:spontaneous,oleg:lebedev,lebedev,romao,pomarol}
 
Whether or not the observed $\cp$ violation in
the kaon sector can arise solely from supersymmetry is a serious
issue. This possibility has been examined in 
\rfs{abel:frere,flavour-structure,cp:only:susy} by simply
assuming $\vckm$ to be real, and subsequently investigating whether
SUSY sources of $\cp$ violation can account for the observed magnitude
of $\epsi_K$ and $\epsi'/\epsi_K$.
Here we would like to draw attention to the particularly attractive SUSY 
scenario with spontaneous $\cp$ symmetry breaking.
The main point of this class of models is that the reality of the 
CKM matrix is automatic (see, e.g., \rfs{gustavo:scpv,bck}), and no longer
an \emph{ad hoc} assumption. 
In the specific scenario we shall
be considering, $\cp$  violation is caused by the phases $\f_D$ and
$\f_N$ associated with the vacuum expectation values of
$\av{{H}_2^0}$ and $\av{N}$ respectively. Being a singlet Higgs field, 
$N$ does not couple to quarks, and although $H^0_2$ 
does couple to the up-type quarks,
the phase appearing in the quark mass matrix can be rotated
away by redefinition of the right-handed quark fields. 
Consequently, this phase does not appear in the CKM matrix since the 
$W$-boson interactions are purely left-handed, but it
does enter in other SUSY interactions.

The plan of this paper is as follows. In \sec{higgs:potential}, we analyse the 
conditions required for the Higgs potential to have a $\cp$-violating 
global minimum.
We show that the upper bound on the mass of 
the lightest Higgs boson imposes constraints on the $\cp$-violating phases, 
$\f_D$ and $\f_N$.  
Section \ref{realCKM} is devoted to a brief review of the 
real CKM matrix in supersymmetric extensions of the SM with two Higgs doublets 
and an arbitrary number of gauge singlet fields. In \sec{nmssm}, we introduce
the mass and mixing matrices, as well as the couplings of the Higgsino, 
$W$-ino, and singlet field in the NMSSM with spontaneous $\cp$ violation.
The calculation of the relevant SUSY contributions to 
$\epsi_K$ in the mass insertion approximation is presented in \sec{scpv}. 
We pay particular attention to the case of low $\tan\b$ and 
examine the question of
whether the choice of parameters of the Higgs 
potential that is required to obtain spontaneous 
$\cp$ violation is consistent 
with the experimentally observed indirect $\cp$ violation in the kaon sector. 
The implications of $\epsi_K$ for the upper bound on the 
lightest Higgs-boson mass and a new flavour structure, besides 
the one of the CKM matrix, are discussed. We also comment on the  
electric dipole moments of electron and neutron. 
Finally, we summarize the key features of our analysis in \sec{conclusions}. 
\section{The Higgs potential}\label{higgs:potential}
\subsection{The superpotential}
We are concerned with the next-to-minimal supersymmetric standard model 
discussed by Davies \ea\ \cite{pot:davies}. The most general 
form of the superpotential that we analyse here 
is\footnote{This form of the 
superpotential has also recently been studied in 
\rf{huber:schmidt}, within the context of electroweak 
baryogenesis.} 
\be\label{pot:eq:W}
W= W_{\subfermion} + W_{\subhiggs},
\ee
with
\be\label{pot:eq:Wfermion}
W_{\subfermion}=\varepsilon_{ab}\left( 
 h_U^{ij}\widehat Q_i^a\widehat U_j\widehat H_2^b 
+h_D^{ij}\widehat Q_i^b\widehat D_j\widehat H_1^a 
+h_E^{ij}\widehat L_i^b\widehat R_j\widehat H_1^a 
\right),
\ee
\be\label{pot:eq:WHiggs}
W_{\subhiggs}= -\lambda \widehat N \varepsilon_{ab} 
\widehat H_1^a\widehat H_2^b -\frac{k}{3}\, {\widehat N}^3 -r \widehat N
-\mu \varepsilon_{ab}  \widehat H_1^a\widehat H_2^b,
\ee
where $i,j=1,2,3$ denote generation indices, $a,b=1,2$ are SU(2) 
indices, and $\varepsilon$ is a completely antisymmetric $2\times2$  
matrix with $\varepsilon_{12}=1$. In the above expression, 
$\widehat H_1^a$ and $\widehat H_2^a$
denote the Higgs doublets of the minimal supersymmetric standard model
and $\widehat N$ is a singlet field. 
The matrices $h_U,h_D$, and $h_E$ give rise to the usual 
Yukawa interactions which generate the masses of quarks and leptons.
Since we are dealing here with spontaneous breaking of $\cp$ rather than 
explicit $\cp$ violation in the Higgs sector, these matrices are real.   
\subsection{Soft-breaking terms}
In addition to the superpotential given by \eq{pot:eq:W}, we
have to specify the SUSY soft-breaking terms. These are given by 
\be\label{pot:eq:soft}
{\mathcal L}_{\sb}={\mathcal L}_{\sb}^{\subfermion}+
{\mathcal L}_{\sb}^{\subhiggs},
\ee
where
\bea\label{pot:eq:softF} 
-{\mathcal L}_{\sb}^{\subfermion}&=& 
{M^2_Q}_{ij}\widetilde Q^{a*}_i\widetilde Q^a_j+{M^2_U}_{ij} 
\widetilde U_i\widetilde U^*_j+{M^2_D}_{ij}\widetilde D_i 
\widetilde D^*_j
+{M^2_L}_{ij}\widetilde L^{a*}_i\widetilde L^a_j\nnu\\ 
&+&{M^2_R}_{ij}\widetilde R_i\widetilde R^*_j
- \frac{1}{2}\left(M_s\lambda_s\lambda_s+M\lambda\lambda 
+M^\prime \lambda^\prime \lambda^\prime+ \hc\right)\nnu\\
&+&\varepsilon_{ab}\left( 
 A_U^{ij}h_U^{ij}\widetilde Q_i^a\widetilde U_j H_2^b 
+A_D^{ij}h_D^{ij}\widetilde Q_i^b\widetilde D_j H_1^a 
+A_E^{ij}h_E^{ij}\widetilde L_i^b\widetilde R_j H_1^a 
\right),  
\eea 
and
\bea\label{pot:eq:softH}
-{\mathcal L}_{\sb}^{\subhiggs} &=&
m_{H_1}^2 H^{a*}_1 H^a_1 + m_{H_2}^2 H^{a*}_2 H^a_2 + m_N^2 N^* N\nnu\\
&-&\left(B\mu \varepsilon_{ab}H_1^a H_2^b +A_{\lambda} N \varepsilon_{ab} 
H_1^a H_2^b +\frac{A_k}{3}N^3 +A_r N + \hc \right).
\eea
\subsection{The scalar potential and spontaneous CP violation}
Following \rf{pot:davies}, we do not require the superpotential
to be invariant under a discrete $Z_3$ symmetry which would imply 
$\mu=r=0$ (thereby solving the so-called `$\m$ problem' of the 
MSSM).\footnote{We remind the reader that a spontaneously broken
discrete symmetry may lead to cosmological domain-wall problems
\cite{domain:walls}. We do not pursue this subject here.} 
Further,  we do not relate the soft SUSY-breaking parameters  
to some common unification scale but rather take them as arbitrary 
at the electroweak scale. 

As it was noted by one of the authors (J.C.R.) a long time ago
\cite{romao}, the NMSSM with a $Z_3$ symmetry has no 
spontaneous $\cp$ violation 
(no-go theorem). The inclusion of the $Z_3$-breaking terms 
in \eq{pot:eq:W}, on the other hand, 
evades that no-go theorem, as was shown by Pomarol \cite{pomarol}. 
Throughout we shall assume that the tree-level 
potential is $\cp$ conserving and take all parameters real, but allow 
complex vacuum expectation values (VEV's) for the neutral Higgs
fields which  emerge after spontaneous symmetry breaking: 
\begin{equation}\label{pot:eq:Hvev}
\vev{H_1^0}=\frac{v_1}{\sqrt{2}}e^{i \theta_1}, \quad
\vev{H_2^0}=\frac{v_2}{\sqrt{2}}e^{i \theta_2}, \quad
\vev{N}=\frac{v_3}{\sqrt{2}}\, e^{i \theta_3}.
\end{equation}
From the superpotential and soft supersymmetry breaking terms, 
Eqs.~(\ref{pot:eq:WHiggs}) and (\ref{pot:eq:softH}), 
we derive the following $\cp$-invariant neutral scalar potential:
\begin{eqnarray}\label{pot:eq:scalarV}
V&=&V_0 +\frac{1}{8}(\lambda_1 v_1^4 +\lambda_2 v_2^4)
+\frac{1}{4}[(\lambda_3 +\lambda_4) v_1^2 v_2^2 +
(\lambda_5 v_1^2 +\lambda_6 v_2^2)v_3^2]\nnu\\
&+&\frac{1}{2}\lambda_7v_1 v_2 v_3^2 
\cos(\theta_1+\theta_2- 2\theta_3)
+\frac{1}{4}\lambda_8 v_3^4+ \frac{1}{\sqrt{2}} \lambda \mu(v_1^2+v_2^2)v_3 \cos(\theta_3)\nnu\\
&+&\frac{1}{2}(m_1^2 v_1^2 +m_2^2 v_2^2 +m_3^2 v_3^2) 
-\frac{1}{\sqrt{2}}m_4 v_1 v_2 v_3 \cos(\theta_1+\theta_2+\theta_3)\nnu\\
&-&\frac{1}{3\sqrt{2}}m_5v_3^3 \cos(3 \theta_3) +m_6^2v_1 v_2
\cos(\theta_1+\theta_2) + m_7^2 v_3^2 \cos(2\theta_3)\nnu\\
&+&\sqrt{2}m_8^3 v_3 \cos(\theta_3),
\end{eqnarray}
where $V_0=r^2$ and
\bea\label{pot:eq:lambdas}
&&\lambda_1=\lambda_2=\frac{1}{4}(g^2+g^{\prime 2}), \quad 
\lambda_3=-\frac{1}{4}(g^2+g^{\prime 2}-4\lambda^2),\quad
\lambda_4=0,\nnu\\
&&\lambda_5=\lambda_6=\lambda^2, \quad
\lambda_7=\lambda k, \quad
\lambda_8=k^2,\\
&&m_{1,2}^2=m_{H_{1,2}}^2 + \mu^2,\quad
m_3^2=m_N^2,\quad
m_4=A_\lambda,\quad
m_5=A_k,\nnu\\
&&m_6^2=-B \mu +\lambda r,\quad
m_7^2=k r,\quad
m_8^3=-A_r.
\eea
In the numerical analysis reported here, we take the $\lambda_l$'s 
($l=1,2,\dots,8$) as given by 
\eq{pot:eq:lambdas}, i.e. in terms of $g$, $g'$, $\lambda$, $k$,  
but consider the $m_l$'s as arbitrary parameters. Referring to 
\eq{pot:eq:scalarV}, we see that the doublet phases enter only in the 
combination $\theta_1+\theta_2$, and thus we may define 
\be\label{pot:eq:phasedef}
\phi_D=\theta_1+\theta_2, \quad \phi_N=\theta_3
\ee
for the doublet and singlet phase respectively. We shall henceforth set 
$\theta_1=0$ in \eq{pot:eq:phasedef}, without loss of generality. 
\subsection{Mass spectrum}
To obtain the physical masses of the NMSSM with SCPV, the following
procedure has been adopted. 
We randomly choose sets of parameters and perform the
minimisation numerically. Next, we check that the local minimum is a
true one not only by verifying the minimisation conditions, but also by ensuring that
all the physical Higgs bosons have positive squared masses. 
In calculating the neutral Higgs-boson mass matrix, radiative
corrections due to top-stop loops have been included.
Finally, we require that the chargino and squark masses are above the
current experimental lower limits \cite{PDG}. 

\begin{figure}
\begin{center}
\psfig{figure=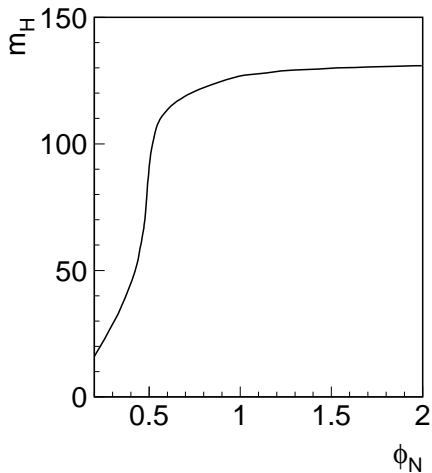,height=2.5in,angle=0}
\vspace{-3mm}
\caption{Maximum value of the lightest Higgs-boson mass (in GeV) as a
function of the $\cp$-violating phase $\phi_N$ (in radians) 
of the Higgs singlet 
field.}\label{scpv_plot_mhtheta} 
\end{center}
\end{figure}

We find that an acceptable mass spectrum can be easily obtained, with  
the exact values depending on the set of parameters we choose. 
Rather than presenting tables or plots for 
different sets of parameters, we will concentrate on two issues: (a)
the influence of the $\cp$-violating phases on the lightest 
Higgs-boson mass; and (b) the maximum theoretically attainable value of
its mass. 
As far as the former is concerned, we confirm the results of an 
analysis of the Higgs potential performed in 
\rf{pot:davies}, where it was pointed out that the large-phase 
solution is favoured. This feature can be clearly observed in 
Fig.~\ref{scpv_plot_mhtheta},  where we display 
the Higgs-boson mass, $m_{H^0}$, as a function of the phase of the gauge 
singlet field.
%
%
It is important to note that large $\cp$ phases may be in 
conflict with existing limits on the electric 
dipole moments (EDM's) of electron and neutron. However, 
these constraints can be evaded, for instance, if 
substantial cancellations among 
different SUSY contributions do occur \cite{edm:canc,edm:analytic} or 
if non-universal soft-breaking terms are present \cite{non-uni}. 
(We will return to this point in a subsequent section.)

\begin{figure}
\begin{center}
\psfig{figure=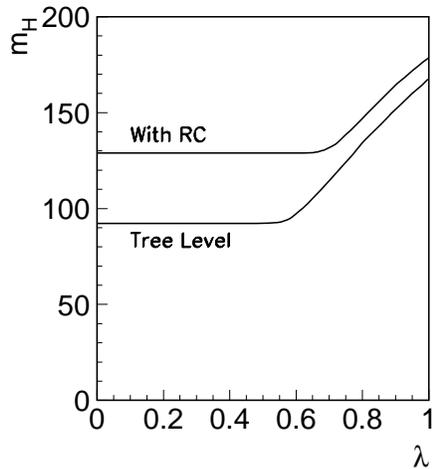,height=2.5in,angle=0}
\vspace{-3mm}
\caption{Maximum value of the lightest Higgs-boson mass (in GeV)
vs $\lambda$ at the tree level and after 
including radiative corrections (at one-loop level) for 
$M_{\susy}=1\,\TeV$.}
\label{scpv_plot_mhlbda} 
\end{center}
\end{figure}

\begin{figure}
\begin{center}
\psfig{figure=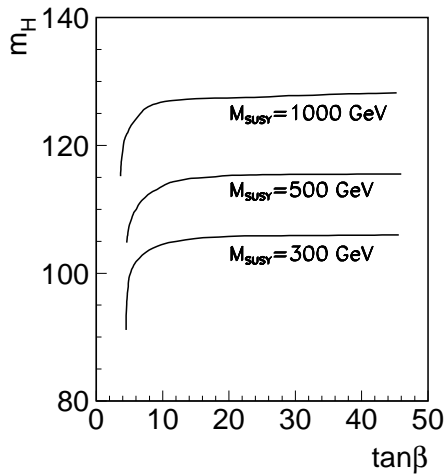,height=2.5in,angle=0}
\vspace{-3mm}
\caption{Upper bound on the lightest Higgs-boson mass (in GeV) 
as a function of $\tan\b$ for various values of $M_{\susy}$.}
\label{scpv_potential_plot} 
\end{center}
\end{figure}

As for the maximal possible value of the Higgs-boson mass, the
result can differ from that of the MSSM for the case of large values of the coupling constant
$\lambda$. In \fig{scpv_plot_mhlbda}, we show the tree level and 
one-loop corrected Higgs-boson mass as a function of $\lambda$.
%
%
It is apparent that only for large values of the
coupling $\lambda$ the situation is different from that of the MSSM. 
For low values of $\lambda$, corrections to the tree level Higgs-boson
mass are significant and depend mainly on the SUSY scale that we take for
the squarks. This is illustrated in  \fig{scpv_potential_plot} where we plot 
the upper limit on $m_{H^0}$ versus $\tan\b\equiv v_2/v_1$
[cf.~\eq{pot:eq:Hvev}] for different values of $M_{\susy}$.
%
%
Before leaving the subject of the Higgs potential, it should be emphasized 
that the Higgs boson 
mass limits obtained at LEP for the standard model and the MSSM (see, e.g., 
\rf{Higgs:mass}) do not necessarily apply to the NMSSM because the 
lightest neutral Higgs boson may have a reduced coupling to the $Z^0$ due to some 
singlet admixture \cite{model:nmssm:I,singlet:admixture}. In this
case, the Higgs boson might not be detectable.
\section{A real CKM matrix}\label{realCKM}
We outline briefly the scenario of a real  CKM matrix in the framework of 
supersymmetry with SCPV. In
this case,  $\cp$ invariance is imposed on the Lagrangian, and
consequently 
all couplings are real. 
To address the question of whether the CKM mechanism occurs in the NMSSM model,
let us consider the quark Yukawa couplings. As mentioned in the
introduction, the gauge singlet field does not couple to quarks. Thus, 
the Lagrangian in terms of the weak eigenstates may be written as 
(omitting generation indices)
\be\label{vckm:eq:lyukawa}
\mathcal{L}_Y = - {h_{U}} {H_{2}^{0*}}\bar{u}_Lu_R -
{h_{D}} {H_{1}^{0*}}\bar{d}_Ld_R + \hc,
\ee
where $h_{U, D}$ are arbitrary real matrices in flavour space. 
After spontaneous symmetry breaking,  the up- and down-type quarks acquire masses, namely
\be\label{vckm:eq:qmass}
m_U = h_U \frac{v_2}{\sqrt{2}} e^{-i\f_D}, \quad m_D= h_D \frac{v_1}{\sqrt{2}}. 
\ee
Note that the overall phase $\f_D$ can be rotated away by means of
a phase transformation on $u_R$,
i.e. $u_R\to u_R' = e^{-i\f_D} u_R$. Since the $W$-boson interactions are purely 
left-handed, this phase does 
not show up in the charged weak  interactions. Consequently,  the CKM matrix is real 
and $\cp$ violation arises solely from the relative phases that appear  in the 
VEV's  of the neutral Higgs fields, \eq{pot:eq:Hvev}.
An alternative way to derive the above result is to compute
the weak basis invariant 
$T\equiv \mathrm{Tr}[H_U,H_D]^3$, where 
$H_U \equiv m_U m_U^\dagger$ and $H_D \equiv m_D m_D^\dagger$ \cite{bernabeu:etal}. 
It follows from \eq{vckm:eq:qmass} that $T$ vanishes, and hence there is no $\cp$ 
violation through  the CKM  mechanism in our model. 
\section{Content of the NMSSM}\label{nmssm}
As we have argued in the previous section,  
all supersymmetry $\cp$-violating phases are 
equal to zero and the CKM matrix is real. 
Even so,  the phases associated with the VEV's,  $\f_{D}$ and $\f_{N}$,  
appear in the scalar quark, gaugino and Higgsino mass matrices, as well as in some of the vertices. We first consider the squark mass matrices. 

\subsection{Squark mass matrices}\label{squark:mass}
The $6\times 6$ squark 
mass-squared matrices in the $(\squarki{L},\squarki{R})$ basis
are given by
\begin{equation}\label{mass:squark:gen}
M^2_{\squark}=
\left(\begin{array}{cc}
M^2_{\squarki{LL}}&M^2_{\squarki{LR}}\\
M^2_{\squarki{RL}} & M^2_{\squarki{RR}}
\end{array}\right),\quad \tilde{q}=\tilde{U}, \tilde{D},
\end{equation}
made up of $3\times 3$ submatrices.
It proves convenient to work in the so-called `super-CKM' basis, in which 
the quark mass matrices are diagonal, and  
squarks as well as quarks 
are rotated simultaneously (see \rf{janusz} for details). 
In this basis, 
the aforementioned submatrices are of the form
\bea\label{mass:squark:detail}
M^2_{\tilde{U}_{LL}}&=& (m_U^{\diag})^2 + V^U_L M_{\tilde{Q}}^2V_L^{U\dagger} +
\frac{1}{6}M_Z^2\cos2\b(3-4\sin^2\theta_W) \unit, \nnu\\
M^2_{\tilde{U}_{RR}}&=& (m_U^{\diag})^2 
+ V_R^U M_{\tilde{U}_{R}}^2V_R^{U\dagger}+
\frac{2}{3}M_Z^2\cos2\b\sin^2\theta_W \unit,\nnu\\ 
M^2_{\tilde{U}_{LR}} &=&M^{2\dagger}_{\tilde{U}_{RL}}=
V_L^U Y_U^*V_R^{U\dagger}\frac{v_2}{\sqrt{2}} 
- \m_{\eff} \cot\b e^{i \f_D} m_U^{\diag},\nnu\\
M^2_{\tilde{D}_{LL}}&=& (m_D^{\diag})^2 +V^D_LM_{\tilde{Q}}^2V_L^{D\dagger}
-\frac{1}{6}M_Z^2\cos2\b(3-2\sin^2\theta_W)\unit, \nnu\\
M^2_{\tilde{D}_{RR}}&=& (m_D^{\diag})^2 + V_R^DM_{\tilde{D}_{R}}^2V_R^{D\dagger}
-\frac{1}{3}M_Z^2\cos2\b\sin^2\theta_W \unit, \nnu\\
M^2_{\tilde{D}_{LR}} &=&M^{2\dagger}_{\tilde{D}_{RL}}=
V_L^D Y_D^*V_R^{D\dagger}\frac{v_1}{\sqrt{2}} 
- \m_{\eff} \tan\b e^{i \f_D} m_D^{\diag},
\eea
$\theta_W$ being the Weinberg angle, $\unit$ denotes a $3\times 3$ unit 
matrix, and the $V$'s diagonalize  the up- and down-type quark mass 
matrices in \eq{vckm:eq:qmass}. [Note that the $V$ matrices are orthogonal, 
and therefore $\vckm={V_L^U} (V_L^D)^{\mathrm{T}}$.] In what follows, we
take a particular case of the `super-CKM' basis where 
$V_L^D=\unit$ and $V_L^U=\vckm$, so that  
\be
(m_U^{\diag})^2 = \vckm (m_U m_U^\dagger) \vckm^{\mathrm{T}}.
\ee
Further, we have introduced the shorthand notation 
\be\label{A-terms}
Y_q^{ij}\equiv A_q^{ij} h_q^{ij}, \quad q=U, D \quad 
(\mathrm{no\ sum\ over\ }i,j), 
\ee
\be\label{eff:mu-term}
\m_{\eff}\equiv \m +\l  \frac{v_3}{\sqrt{2}}e^{i\f_N}.
\ee
[Notice that \eqs{mass:squark:detail}{eff:mu-term} show explicitly the 
dependence on the $\cp$ phases.] 
As we shall see in the next section, a 
non-universal flavour structure 
in the 
$A$ terms, i.e. $A^{ij}_q\neq \mathrm{constant}$, is indispensable
for having sizable supersymmetry contributions to $\cp$ violation in the 
kaon sector.\footnote{For related work, see 
\rfs{abel:frere,flavour-structure}.}  
At this stage, we therefore make no assumptions
regarding the nature of the several matrices and 
couplings  involved. 

The squark mass-squared matrix, $M^2_{\squark}$, can then be diagonalized by 
a $6\times 6$ unitary matrix such that
\begin{equation}\label{mass:squark:diag}
(M_{\squark}^{\diag})^2 = R_{\squark} M_{\squark}^2
R_{\squark}^{\dagger}. 
\end{equation}
%
\subsection{Chargino mass matrix}\label{chargino:mass}
We now turn our attention to the four-component Dirac fermions, 
$\chargino_{1,2}$, which arise from the mixing of 
the $W$-inos and the charged Higgsinos. Within SUSY, the chargino mass
terms in the Lagrangian are
\begin{equation}\label{mass:charg:gen}
\mathcal{L}_{\mathrm{mass}} = -\frac{1}{2} (\psi^{+T}, \psi^{-T})\left(
\begin{array}{cc}
0 & M_{\C}^T\\
M_{\C} & 0
\end{array} 
\right)\left(
\begin{array}{cc}
\psi^+\\
\psi^-
\end{array}\right)+ \hc,
\end{equation}
where $\psi^{+T}=(-i\l^+, \tilde{H}^+_2)$, $\psi^{-T}=(-i\l^-,\tilde{H}^-_1)$.
The chargino mass matrix in the NMSSM with spontaneous 
$\cp$ violation reads 
\begin{equation}\label{mass:charg:matrix}
M_{\C}= \left(\begin{array}{cc}
M_2 & \sqrt{2}M_W \sin\b e^{-i \f_D}\\ \sqrt{2}M_W\cos\b 
&\m_{\eff}\end{array}\right)
\end{equation}
with the gaugino mass $M_2$ being real and $\m_{\eff}$ given in 
\eq{eff:mu-term}.
As usual,  $M_{\C}$ can be diagonalized by means of 
a biunitary transformation, i.e.
\begin{equation}\label{mass:charg:diag}
M^{\diag}_{\C}= U^* M_{\C}V^\dagger,
\end{equation}
where $M^{\diag}_{\C}$ is positive and diagonal, and $U$, $V$
are unitary matrices.

For convenience, we perform our calculation in the weak (rather than
the physical) basis, i.e.~$\tilde{W}^-=(-i\l^-,i\ol{\l}^+)^{\mathrm{T}}$,
$\tilde{H}^-=(\tilde{H}_1^-,\ol{\tilde{H}}_2^+)^{\mathrm{T}}$,
and their relation can be summarized as 
follows:\footnote{We define $\tilde{W}^-$ and $\tilde{H}^-$ as particles, contrary to 
\rf{susy:rev:haber}.}
\begin{subequations}\label{box:eq:wino chargino}
\be
P_L \wino^{-}= P_L (U_{11}^{*} {\tilde{\chi}}^{-}_{1}
+U_{21}^{*} {\tilde{\chi}}^{-}_{2}),\quad
P_R \wino^{-}= P_R (V_{11} {\tilde{\chi}}^{-}_{1} 
+V_{21} {\tilde{\chi}}^{-}_2),
\ee
\be
P_L \higgsino^{-}=  P_L (U_{12}^{*} {\tilde{\chi}}^{-}_{1}
+U_{22}^{*} {\tilde{\chi}}^{-}_{2}),\quad
P_R \higgsino^{-}=P_R (V_{12} {\tilde{\chi}}^{-}_{1}
+V_{22} {\tilde{\chi}}^{-}_{2}),
\ee
\end{subequations}
with $P_{L,R}=(1\mp \g_5)/2$. In order to specify the relevant couplings 
within the NMSSM, we recall that \cite{susy:rev:haber} 
\be\label{L:int:charg}
-\mathcal{L}^{\tilde{W}, \tilde{H}}= 
g({H^0_1}^* \overline{\tilde{W}} P_L \tilde{H} +
{H^0_2}^* \overline{\tilde{H}} P_L \tilde{W})
+\l N \overline{\tilde{H}} P_L \tilde{H} + \hc
\ee
Then, substituting the VEV's from \eq{pot:eq:Hvev},
and adding $W$-ino and Higgsino `mass' terms,  we arrive at
\bea\label{int:lagrangian}
-{\mathcal L}_{\mathrm{int}}= m_{ \tilde{W}} \overline{\tilde{W}}\tilde{W}
+  m_{ \tilde{H}} \overline{\tilde{H}} \tilde{H}
+\frac{g}{\sqrt{2}}(v_1 e^{-i \varphi} \overline{\tilde{W}}_R\tilde{H}_L+
v_2 e^{i \f_D} \overline{\tilde{W}}_L\tilde{H}_R+ \hc),\hspace{-2em}
\nnu\\
\eea
where
\be\label{wino:higgsino}
m_{ \tilde{W}}=M_2, \quad 
m_{ \tilde{H}}=  |\m_{\eff}|, \quad
\varphi= \arg \left( \m_{\eff}\right),
\ee
and $\m_{\eff}$ as in \eq{eff:mu-term}. In deriving  \eq{int:lagrangian},
we have adjusted our definition of 
the phase of the left-handed Higgsino field such 
that  $\tilde{H}_L\to e^{-i \varphi}\tilde{H}_L$.
As for gluino and neutralino interactions, we will argue
in the next section that to good approximation the chargino box diagram may 
be regarded as the dominant contribution to $\epsi_K$
(see also \rfs{lebedev,pomarol}). 
\section{Implications of indirect CP violation for the NMSSM}\label{scpv}
In this section, we wish to explore the consequences  of SCPV for the upper bound of the lightest 
Higgs-boson mass derived in \sec{higgs:potential} by taking into account
$\cp$ violation in $K^0$--$\bar{K}^0$ mixing. To accomplish this, 
we will compute the box-diagram 
contributions to $\epsi_K$ by applying the mass insertion
approximation \cite{mia}. That is, 
adopting the notation 
\be
M_{\squark}^2\equiv (M_{\squark}^0)^2+{M_{\squark}'}^2, \quad 
(M_{\squark}^0)^2=\diag(a_1^0,\dots, a_6^0), 
\ee
and  recalling \eq{mass:squark:diag}, we may write to first order in ${M_{\squark}'}^2$ \cite{mia:buras:etal}
\bea
(R^\dagger_{\squark})_{mk} f(a_k) (R^{}_{\squark})_{kn}
 = \d_{mn}f(a_n^0) + 
({M_{\squark}'}^2)_{mn} F(a_m^0,a_n^0), 
\eea 
$f$ being an arbitrary function, $k=1,\dots,6$, and
\be
F(x,y)=\frac{f(x)-f(y)}{x-y}.
\ee
\subsection{Effective Hamiltonian}
Let us start with the effective Hamiltonian governing $\D S=2$ transitions, 
which can be written in the form \cite{eff:review}:
\be\label{heff}
\heff= \sum_i c_i {\mathcal O}_i,
\ee
where the $c_i$'s are the Wilson coefficients describing the short-distance
interactions while ${\mathcal O}_i$ denote local operators which may
be found, e.g., in  \rfs{fcnc:susy:constraints,deltas2:operators}.
The off-diagonal element of the neutral kaon mass matrix is related to $\heff$ through
\be\label{mdef}
\mathcal{M}_{12}=\frac{\braket{K^0}{\heff}{\bar{K}^0}}{2m_K},
\ee
and its imaginary part gives rise to (assuming $\epsi' \ll \epsi_K$) 
\be\label{eK:eq:def}
\epsi_K \simeq \frac{e^{i\pi/4}}{\sqrt{2}}
\frac{\Im\mathcal{M}_{12}}{\Delta m_K},
\ee
with the experimental values of 
$\D m_K = (3.489\pm0.008)\times 10^{-12}\ \MeV$ 
\footnote{Note that for $\D m_K$, and following standard procedure, we
  have used the experimental value. No useful bound on the present
  model can be derived from $\left.\D m_K\right|_{\mathrm{exp}}$, 
since it receives
  important long-distance contributions.}
and $|\epsi_K| = (2.271\pm 0.017)\times 10^{-3}$ \cite{PDG}.

The Wilson coefficients $c_i$ in the presence of SUSY contributions 
can be decomposed as follows: 
\be\label{cis}
c_i= c_i^W + c_i^{\higgs}+ c_i^{\chargino} + c_i^{\gluino} + 
c_i^{\neutralino} .
\ee
%
%
\begin{figure}
\begin{center}
\epsfig{figure=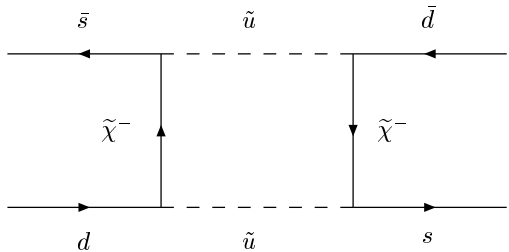,height=1.3in,angle=0}
\caption{The dominant box-diagram contribution to the off-diagonal element 
$\Im {\mathcal M}_{12}$ in the neutral kaon mass matrix within the framework of 
the NMSSM.}\label{fig:box:full} 
\end{center}
\end{figure}
Regarding the various contributions in \eq{cis}, we confine ourselves to a
relatively few remarks.
First, there are no $W$-boson and charged Higgs-boson 
contributions to the imaginary part of $\mathcal M_{12}$ since the CKM matrix is real.
Second, in subsequent calculations we will choose a basis where the down-type 
Yukawa matrices are diagonal. 
As a consequence, the flavour off-diagonal gluino contributions 
are either zero or real in the approximation of retaining only a 
single mass insertion in an internal squark line.
Third,  box diagrams with neutralinos are expected to compete with the 
chargino contributions only for large values of $\tan\b$ 
(i.e. $\tan\b \sim 50$), and hence are less important for the present scenario with low $\tan\b$. 
In the remainder of this section, we will therefore focus on the 
chargino box diagram with scalar up-type quarks (see \fig{fig:box:full})
as it gives by far the dominant contribution in our model.

Turning to the operators ${\mathcal O}_i$ in \eq{heff}, the $\D S=2$ 
transition within the model under study  
is largely governed by the $V$--$A$ four-fermion operator 
${\mathcal O}_1 = \overline{d} \g^\m P_L s \overline{d} \g_\m P_L s$. 
As a matter of fact,  the new-physics contribution to the 
remaining operators (i.e. scalar and pseudoscalar operators) are suppressed 
relative to the chargino contribution
either through small quark masses, $m_{s,d}$, or small Yukawa couplings. 
Thus, it is fairly reasonable to consider only the 
non-standard contribution to 
the Wilson coefficient $c_1$. 
The relevant hadronic matrix element in the 
vacuum saturation approximation is given by 
$\braket{K^0}{{\mathcal O}_1}{\bar{K}^0}= (2/3) f_K^2m_K^2$, 
with $f_K=160\ \MeV$.  
\subsection{Chargino box diagram}
\begin{figure}
\begin{center}
\epsfig{figure=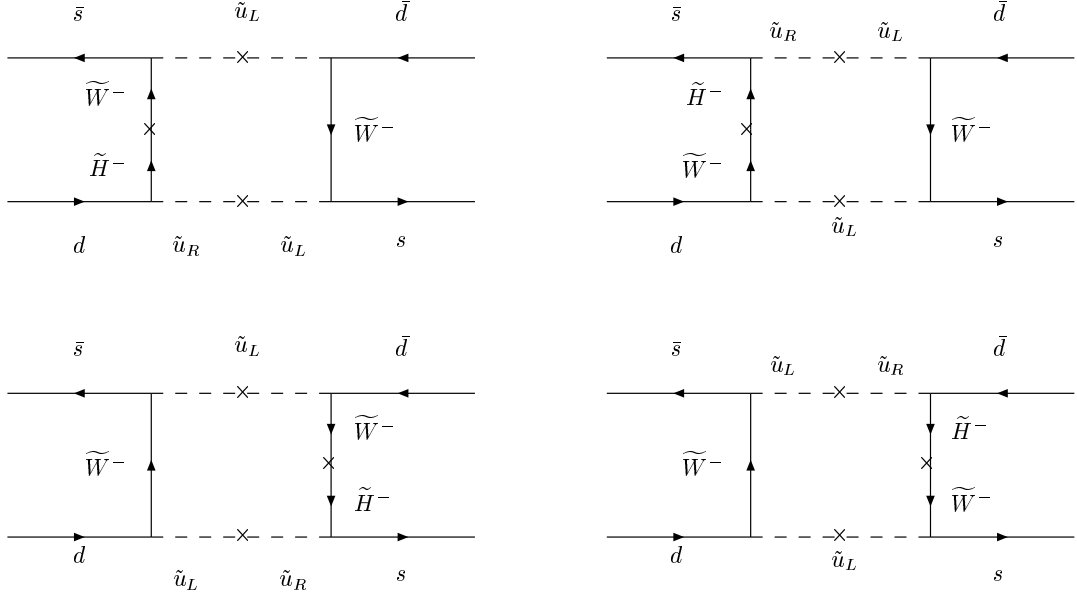,angle=0,height=3.2in}
\caption{The main contributions to $\epsi_K$ in the mass insertion approximation
with $W$-ino and Higgsino exchange.}\label{mia:leading} 
\end{center}
\end{figure}

In the $\tilde{W}$--$\tilde{H}$ basis, and making use of the mass
insertion approximation, the chargino contributions to the short-distance 
coefficient $c_1$ can be written as a sum of 16 individual terms, each
corresponding to a particular box diagram. 
Introducing the notation $\{ij\}$ $(i,j=L,R)$ for the various mass insertions, 
these contributions can be
classified in terms of the nature of the mass insertion in the internal
squark line and the number of $\tilde{W}$--$\tilde{H}$ exchanges in the
sfermion line.~That is, the relevant amplitude for $c_1$ 
can be  symbolically written as
\bea
A_{c_1}^{\chargino} &\sim&  
(X_{\tilde{W}\mbox{--}\tilde{H}})^0[\{LL\}\{LL\} + \{RR\}\{RR\} + \{LR\}\{RL\}] \nnu \\
&+& (X_{\tilde{W}\mbox{--}\tilde{H}})^1 
[\{LL\}\{LR\} + \{LR\}\{RR\}] \nnu \\
&+& (X_{\tilde{W}\mbox{--}\tilde{H}})^2[\{LL\}\{RR\} + \{LR\}\{LR\}].
\eea
Only one diagram belongs to the $\{LL\}\{LL\}$ and $\{RR\}\{RR\}$ mass 
insertions, whereas we
find two (four) diagrams associated with the $\{LR\}\{RL\}$, 
$\{LL\}\{RR\}$, $\{LR\}\{LR\}$ ($\{LL\}\{LR\}$, $\{LR\}\{RR\}$) 
mass insertions.
It turns out that the dominant contribution to $c_1$
corresponds to the class of diagrams with $(X_{\tilde{W}\mbox{--}\tilde{H}})^1 
\{LL\}\{LR\}$, depicted in \fig{mia:leading}, in agreement with the findings 
of \rfs{lebedev,pomarol}. 
%
%
%
The calculation of the dominant contributions to
$\Im \mathcal{M}_{12}$ for the case of non-degenerate left-handed 
up-squarks leads to a cumbersome expression. Hence, for the sake of 
simplicity, we perform the calculation in the 
limit of degenerate left-handed up-type squarks, and find that the
expression for $\Im \mathcal{M}_{12}$ becomes much simpler, and its 
physical interpretation more transparent.
(We have numerically verified that small values of the off-diagonal
elements of ${M^2_{\tilde{Q}}}$ have little effect on the 
$\tilde{u}_L$ mass splitting.)
In that limit, i.e.~$m_{\tilde{u}_L}=\av{m_{\tilde{q}}}$, where $\av{m_{\tilde{q}}}$ 
represents the average squark mass of the first two generations,
we obtain
\bea\label{eK:eq:s4s13}
&&\Im \mathcal{M}_{12}=
\frac{2 G_F^2 f^2_K m_K m_W^4 }{3\pi^2\av{m_{\tilde{q}}}^8}
(V_{td}^*V_{ts}^{}) m_t^2 
\left|e^{i \phi_D}m_{{\wino}}+\cot\b m_{{\higgsino}}\right| \nnu\\ 
&&\times\left\{\D A_U \sin(\varphi_{\chi}-\f_D)\;
{(M^2_{\tilde{Q}})}_{12}\;
I(r_{\tilde{W}},r_{\tilde{H}},r_{\tilde{u}_L}, r_{\tilde{t}_R})\right\},
\eea
with $m_{\tilde{H},\tilde{W}}$ and ${(M^2_{\tilde{Q}})}_{12}$
as defined in Eqs.~(\ref{wino:higgsino}) and (\ref{mass:squark:detail}) 
respectively. Here we have used unitarity of the CKM matrix and 
kept only the leading top-quark contribution. In the above formula,  
$r_i=m_i^2/\av{m_{\tilde{q}}}^2$, $\D A_U\equiv A_U^{13}-A_U^{23}$, and 
\be
\varphi_{\chi} = \arg \left(e^{i \phi_D}m_{{\wino}}
+\cot\b m_{{\higgsino}}\right).
\ee
The function $I$ in \eq{eK:eq:s4s13} can be reduced successively to 
elementary functions for appropriate limits of its arguments
(see \rfs{edm:analytic,mia:buras:etal}). For ease of writing, 
we give the result for the special case
\be
I(x,x,1,1)=\frac{1}{2}\left[
\frac{10+19x+x^2}{3(1-x)^5}+\frac{(1+6x+3x^2)}{(1-x)^6}\ln x\right],
\ee
which corresponds to the scenario with $m_{\tilde{H}}=m_{\tilde{W}}$ and 
degenerate squarks. The appearance of $A_U^{i3}$ ($i=1,2$) terms in 
\eq{eK:eq:s4s13} is
closely related to having the Higgsino coupling to down and strange
quarks in distinct diagrams, and stems from the trilinear soft-breaking 
terms in $Y_U^{ij}\equiv A_U^{ij} h_U^{ij}$ [\eq{A-terms}].
From inspection of \eq{eK:eq:s4s13}, it is 
straightforward to conclude that in order to get 
$\Im \mathcal{M}_{12}$ different from 
zero we need to move aside from a theory of universal $A_U$ terms
(i.e. $\D A_U\neq 0$) -- in other words, it is not possible to saturate the 
observed $\cp$ violation in the $K$-meson system in the context of SUSY 
with a real CKM matrix and universal $A_U$ terms 
\cite{abel:frere,flavour-structure}.\footnote{It has 
recently been pointed out in \rf{demir} 
that there might be a connection between a non-trivial flavour structure and 
sizable SUSY  $\cp$ violation, thereby avoiding domain-wall problems.} 
Let us note parenthetically that the need for a special $A_U^{i3}$
($i=1,2$) texture as a key element to get the experimentally measured 
$\epsi_K$ has also been  pointed out in \rf{abel:frere}, 
although in a different scenario.

Since the sign of $\D A_U$ is not fixed, we do not 
include the constraint $\Re \epsi_K > 0$ in the analysis which follows.
\subsection{Numerical results and discussion}\label{res:disc}
For our numerical calculations, we have used the nominal values
\be
{(M^2_{\tilde{Q}})}_{12}/\av{m_{\tilde{q}}}^2=0.08 ,\quad 
V_{td}=0.0066, \quad V_{ts}=-0.04, \quad m_t=175\,\GeV.
\ee 
Notice that the value of $V_{td}$ differs from the one determined 
in the context of the SM with CP violation by fitting
to experimental data; 
this is due to the fact that we are dealing 
with a flat triangle and the corresponding orthogonality constraints.\footnote{The value used for $|V_{td}|$ is consistent with the 
experimental value of $B_d^0$--$\bar{B}^0_d$ mixing, provided one assumes a
new-physics contribution to $\D m_{B_d}$ of at least $20\%$. To be specific, 
we have chosen a positive value for $\rho$. As for a negative value of $\rho$,
orthogonality of the CKM matrix would imply $|V_{td}|=0.011$.}
The remaining parameters are  evaluated  numerically by minimising the 
Higgs potential (see \sec{higgs:potential}).
Our results for the absolute value of 
$\epsi_K$ for various sets of SUSY parameters and low $\tan\b$
are reported in Table \ref{table:res}. Recall that the numerical 
values we present for $|\epsi_K|$ have been derived for 
degenerate left-handed squarks. 
Taking into account corrections due to non-degeneracy will lead to a
very small enhancement, typically a few percent, 
for $\Im {\mathcal M}_{12}$.
Hence, the 
values for $|\epsi_K|$ listed in Table \ref{table:res} may be interpreted 
as a lower bound.
%
%
\begin{table}
\caption{Numerical values of $|\epsi_K|$ in the low $\tan\b$ region
for certain sets of model parameters that satisfy the minimisation 
condition of the Higgs potential. We have chosen $\D A_{U}=500\,\GeV$ for 
the non-universal $A$ terms, as described in the text.}\label{table:res}
\vskip 5pt
\tabcolsep=7.9pt
\begin{center}
\begin{tabular}{cccccccccc} \hline\hline
$|\epsi_K|$ &$\phi_D$  &$\phi_N$ & 
$m_{H^0}$ & $\av{m_{\tilde{q}}}$ &
$m_{\tilde{t}_R}$ &$\tan \beta$  &$\lambda$ & $v_3$\\
$(10^{-3})$&$(\mathrm{rad})$ & $(\mathrm{rad})$ &$(\GeV)$&$(\GeV)$&$(\GeV)$&
&&$(\GeV)$ \\ \hline 
$3.24$ &$4.71$   &$1.57$  &$99$ &$252$ &$235$ &$6.7$&$-0.03$ &$327 $ \\ 
$3.03$ &$0.89$   &$1.75$   &$97$ &$261$ &$168$ &$6.6$&$+0.33$ &$387$ \\ 
$2.75$ &$4.71$   &$4.71$   &$99$ &$232$ &$201$ &$9.2$&$-0.02$ &$221$ \\ 
$2.42$ &$1.96$   &$4.08$   &$94$ &$299$ &$174$ &$5.1$&$-0.06$ &$352$ \\ 
$2.10$ &$4.67$   &$4.75$   &$98$ &$279$ &$220$ &$7.8$&$+0.01$ &$142$ \\ 
$2.02$ &$4.68$   &$4.71$   &$92$ &$250$ &$152$ &$7.4$&$+0.02$ &$371$\\ 
$2.01$ &$4.18$   &$4.73$   &$96$ &$280$ &$232$ &$4.6$&$-0.01$ &$238$ \\
$1.31$ &$1.12$   &$4.72$   &$100 $&$273$&$241$ &$9.6$&$-0.01$ &$238$ \\ 
$1.29$ &$2.35$   &$4.70$   &$99$ &$258$&$230$ &$6.1$&$-0.13$ &$363$ \\ 
\hline\hline
\end{tabular}
\end{center}
\end{table}
As far as $M_2$ and $\mu$ are concerned, we obtain the ranges 
$90\,\GeV \lesssim M_2 \lesssim 160\, \GeV$ and 
$120\,\GeV \lesssim |\m| \lesssim 270\, \GeV$.
Turning to the non-universal flavour structure,
it is obvious  from 
\eq{eK:eq:s4s13} that there is a linear dependence of $\epsi_K$ on the 
relative difference $\D  A_{U}$. In order to saturate the value of 
$|\epsi_K|$ and to obey present experimental 
limits on the sparticle spectrum, one has to take $\D A_{U}$ of order 
$500\,\GeV$.
A detailed discussion of scenarios where such values for 
$A_U^{i3}$ ($i=1,2$)  appear lies beyond the scope of this 
paper. We will just mention that values around the $\TeV$ scale do not 
significantly affect the mass spectrum of the theory, and that they can 
account for values of the left-right mass insertion, i.e.
\be
(\d_{LR}^U)_{i3}=\frac{v_2}{\sqrt{2}}\,\frac{(V_L^U Y_U^*V_R^{U\dagger})_{i3}}{\av{m_{\tilde{q}}}^2},  
\ee
as small as $0.02$ and $0.08$ for $(\d_{LR}^U)_{13}$ and 
$(\d_{LR}^U)_{23}$ respectively, which 
are consistent with the bounds  
coming from the measurement of $b\to s \g$ branching fraction. (Note that 
these $\d$'s also affect the decay $b\to s l^+l^-$.) Indeed, according to 
\rf{janusz}, $(\d_{LR}^U)_{23}$ is constrained 
to be $|(\d_{LR}^U)_{23}| \lesssim 3 (\av{m_{\tilde{q}}}/500\,\GeV)^2$,
which is only useful if $\av{m_{\tilde{q}}} \lesssim 300\, \GeV$.
Similarly, the chargino contribution to $B^0_d$--$\bar{B}^0_d$ mixing  
leads to the constraint 
$(\d_{LR}^U)_{13} \lesssim 0.1 (\av{m_{\tilde{q}}}/500\,\GeV)$ 
\cite{mia:buras:etal}.\footnote{Note 
that the requirement of vacuum  stability also leads to upper bounds on 
$(\d_{LR}^U)_{i3}$ ($i=1,2$) \cite{ccb}. The nominal values used in our numerical analysis are compatible with these upper limits.}

At this point, a few remarks are in order regarding previous studies
contained in the literature. 
As we have already mentioned, the diagrams depicted in \fig{mia:leading} 
give the leading contributions to $\Im \mathcal{M}_{12}$, in agreement with 
\rfs{lebedev,pomarol}.
The final form of our result, however, differs from that of previous works 
in several ways. First, we have used the mass insertion approximation
to compute the dominant contributions to $\epsi_K$, which enables us to use 
other largely model-independent constraints on the off-diagonal elements of 
the squark mass matrices.
Second, we have considered a non-trivial structure for the $LR$ squark  
mass-squared 
terms, which proved to be crucial in order to have indirect CP
violation within the present model. More importantly, our parameter space
is severely constrained since in the calculation of the various
contributions to $\epsi_K$ we have considered only those sets of parameters
that correspond to true minima of the Higgs potential. 

We conclude this section with several comments on large $\cp$ phases 
and their implications for the EDM's of electron and neutron.

\subsection{Remarks on the electric dipole moments} 
As seen from Table \ref{table:res},
the $\cp$ phase of the gauge singlet field, $\phi_N$, is accompanied by 
small values of $\l$ so that the EDM constraints are less 
stringent \cite{edm:lambda}. 
Further, the SUSY contributions to the 
EDM's depend only on flavour-diagonal entries in the squark mass matrix,  
whereas the expression for $\epsi_K$ involves flavour off-diagonal elements.
However, it is evident from \eq{mass:squark:detail} that there is an 
overall $\cp$-violating phase which is independent of the 
new flavour structure.
Apart from the requirement that $A_U^{ij}|_{i=j}\ll A_U^{ij}|_{i\neq j}$,
a conceivable way to suppress the EDM's
includes large cancellations between different SUSY 
contributions and a SUSY particle spectrum with masses of scalar quarks and 
gauginos in the $\TeV$ range. 
Given the analytic results for the contributions to the EDM's of electron and 
neutron mediated by photino and gluino 
\cite{fcnc:susy:constraints}, together with the sets of parameters 
displayed in Table \ref{table:res} and the present experimental results of 
$d_n < 6.3\times 10^{-26}\, e\, \mathrm{cm}$ ($\cl{90}$) and
$d_e = 1.8\times 10^{-27}\, e\, \mathrm{cm}$ \cite{PDG},
the photino and gluino masses are 
required to satisfy  
$0.5 \, \TeV \lesssim m_{\tilde{\g}} \lesssim 2\, \TeV$ and
$2  \, \TeV \lesssim m_{\gluino} \lesssim 6\, \TeV$.\footnote{For simplicity, 
we have assumed $m_{\tilde{l}}=m_{\squark}$. Note that the 
constraint on $m_{\gluino}$ is  relaxed if we allow for an order-of-magnitude
variation in the value of the hadronic matrix elements.}

Such a hierarchy in the soft gaugino masses requires further
discussion. We first note that masses of the superpartners of about $1\, \TeV$ 
may be in conflict with the cosmological relic density. In addition,
such a scale for the $\tilde{\gamma}$ and the $\tilde{g}$ masses is
rather unnatural, since the masses of the squarks and $W$-ino 
are typically of the order $100$--$300\, \GeV$ in the model 
under consideration.
[Similar conclusions can be drawn if we take into account
the chargino and neutralino contributions (see, e.g., \rf{edm:analytic}).]
Finally, we stress
that the above-mentioned hierarchy for the scalar partners leads to an
unacceptable scenario for the lightest supersymmetric particle (LSP). In this 
case, the LSP would be either charged (one of the lightest squarks or a 
chargino), or would have a non-zero lepton number.
To summarize, it appears that unless the parameters of the model are 
fine-tuned, there will be large contributions to the EDM's of 
electron and neutron, in potential conflict with experiment.\footnote{As far 
as the $B$ system is concerned, we merely remark that the model can 
accommodate a large CP asymmetry $a_{J/\psi}$, i.e.~a large value of 
$\sin2\b$ (see, e.g., \rf {oleg:lebedev}).}
\section{Conclusions}\label{conclusions}
The origin of $\cp$ violation, in particular, the question of whether 
$\cp$ is explicitly or spontaneously broken, is an important issue in particle physics. In this paper, we have studied spontaneous $\cp$ violation in the 
context of the NMSSM, demonstrating that it is possible to generate sufficient 
$\cp$ violation in order to account for the magnitude of $\epsi_K$. We have 
emphasized that the NMSSM with spontaneous $\cp$ violation  is a natural 
framework to  discuss the possibility of having $\cp$ violation that arises
entirely from SUSY interactions. This is due to the fact that in the 
above-mentioned scenario the CKM matrix is automatically real. 

By performing a complete and systematic study of the Higgs
potential in the NMSSM with spontaneous $\cp$ violation,
we have shown that the minimisation of the most general 
Higgs potential leads to an acceptable mass spectrum which is accompanied by
large $\cp$-violating phases. We have 
argued that the lightest neutral Higgs boson 
can have a reduced coupling to the $Z^0$ due to the additional singlet field, 
and thus may escape detection at LEP. We have shown that it is possible to 
have a mass of the lightest neutral Higgs boson of about $125\,\GeV$. 
However, such values for the 
Higgs-boson mass require a relatively large value for $M_{\susy}$ of the 
order $O(1\,\TeV)$. By contrast, a rather low SUSY scale with 
$M_{\susy}\approx 300\,\GeV$ (i.e.~light squark and $W$-ino masses) 
and a non-trivial flavour structure of the soft SUSY-breaking trilinear 
couplings $A_U^{i3}$ ($i=1,2$) are required in order to account for the  
observed $\cp$ violation in $K^0$--$\bar{K}^0$ mixing. 
As a consequence, the parameter space is severely constrained and the mass of 
the lightest Higgs boson is further diminished,
and it turns out to be no greater than $\sim 100\,\GeV$ for the case of low $\tan\b$ ($\lesssim 10$).
As far as large $\cp$ phases are concerned, we have argued that 
it may be difficult to reconcile the large-phase solution
with the severe constraints on the EDM's of electron and neutron.
Although we do not exclude the possibility of cancellations between different 
contributions, we find this highly unlikely in view of the 
constrained parameter space. Therefore, the implications of the 
EDM bounds on the parameter space, as well as the implied LSP
scenario, are a great challenge for SUSY models with spontaneous CP
violation (at least within the minimal model we have considered here).

\acknowledgments
One of us (A.M.T.) would like to thank Ricardo Gonz\'alez Felipe 
for useful discussions, and 
the Theory Division at CERN for its kind hospitality during the final 
stage of this work. This project was supported in part by the 
TMR Network of the EC under contract ERBFMRX-CT96-0090. 
A.M.T. acknowledges support by  
`Funda\c c\~ao para a Ci\^encia e Tecnologia' under grant PRAXIS XXI 
BD/11030/97. 

%
%
 
%
\end{document}